\documentstyle[aps,prl,epsfig,twocolumn]{revtex}
 \begin{document}
\twocolumn[\hsize\textwidth\columnwidth\hsize\csname 
@twocolumnfalse\endcsname
\draft

\title{$p$-wave Cooper Pairing of Fermions in  Mixtures of Dilute
Fermi and Bose Gases}

\author{D.~V.~Efremov $^{1,2}$ and L. Viverit $^{3}$}
\address{$^{1}$Max-Planck-Institut f\"ur Physik Komplexer Systeme, 
N\"othnitzer Str. 38, 01187 Dresden, Germany 
}
\address{$^{2}$Kapitza Institute for Physical Problems, 117334 Moscow, Russia}
\address{$^{3}$Dipartimento di Fisica, Universit\`a di Milano, via Celoria 16,
20133 Milano, Italia}
 
\maketitle

\widetext

\begin{abstract}

We predict a $p$-wave  
Cooper pairing of the spin-polarized fermions in a binary 
fermion-boson mixture  
due to the exchange of density fluctuations of the bosonic medium. 
We then examine the dependence of the Cooper paring temperature
on the parameters of the system. 
We finally estimate the effect
of combining the boson-induced interaction with other pairing
mechanisms, e.g the Kohn-Luttinger one, 
and find that the critical temperature of $p$-wave Cooper pairing can be
realistic for experiment.     

\end{abstract}

]
\narrowtext

The investigation of quantum degenerate gases has a long history.          
The latest step ahead in this direction has been the realization of 
Bose-Einstein 
condensation in gases of the alkali elements $^{87}$Rb \cite{Anderson95},
$^{7}$Li \cite{Bradley95}, and $^{23}$Na \cite{Davis95}
in the confined geometry of optomagnetic traps.
Currently experimentalists are concerned with obtaining 
fermionic superfluidity.
The regime of quantum degeneracy has been already achieved experimentally 
in samples of potassium
\cite{Marco99} and lithium atoms \cite{Schreck00}. 
By the first group degeneracy has been reached by evaporatively cooling 
together two hyperfine states of the same
fermionic element ($^{40}$K) and by the second group by sympathetically 
cooling the fermionic isotope $^6$Li with the bosonic one $^7$Li.
Other groups are progressing along similar lines.

Various articles appeared lately discussing 
what type of Cooper pairing ($s$-wave or $p$-wave)
is more realistic to achieve \cite{Stoof00}
-\cite{Efremov00a}. 
From one side $s$-wave pairing 
(with orbital angular momentum of the pair $L=0$) 
has a higher critical temperature than $p$-wave ($L=1$) for typical 
parameters.
However it takes place only between atoms of different spins  
(hyperfine components).
This imposes a very stringent 
constraint  on the densities $n_1$, $n_2$ of the components which form 
the pairs. One should have
$\left|( n_1-n_2)/(n_1 + n_2) \right|
\leq T_c/\varepsilon_F \ll 1$.
(Here and below we let $k_{B}=\hbar=1$).
From the experimental point of view, with alkalis
this condition may be hard to achieve as the two hyperfine components
are manipulated independently.
Since on the other hand a $p$-wave Cooper pair is formed by atoms in
the same spin component, the restriction on the densities 
is taken away, and it is therefore relevant to ask how realistic it
would be to consider observing that instead.

As a contribution to this field, in this letter we want 
to answer the question of what is the effect of 
the presence of a Bose gas on the $p$-wave Cooper pairing of fermions. 
This question is also
very important for understanding the role of phonons in nontrivial 
superconductivity in such materials as  HTSC, heavy-fermon systems,
organic superconductors, and Sr$_2$RuO$_4$ which present superfluidity
with non-zero orbital angular momentum of the Cooper 
pairs \cite{Abrikosov,Kohmoto}.     
Low densities of atomic gases give the possibility to introduce small
parameters in the theory -- namely the gas parameters  
$(a n^{1/3}) \ll 1$,  where $a$ -- stands for the appropriate
two particle scattering length in vacuo and $n$ 
for the density of the Bose or the Fermi gas. 
Using this small
parameters we calculate the effective interaction. 
We show that the exchange of boson density fluctuations 
gives an attractive contribution 
to the effective interaction of two fermions with bosonic superfluid
background in channels with non-zero angular momentum. The largest one
corresponds to the $p$-wave channel.
Hence the fermions in binary mixtures of bose and fermi gases are unstable
towards $p$-wave Cooper pairing. We calculate the associated $T_c$, and
determine the Fermi and 
Bose densities which provide the highest value of $T_c$
compatible with the constraints imposed by the instability to phase 
separation. 
At the end of the paper we show that the boson-induced interaction can be
combined with some other $p$-pairing mechanism, in which case it 
acts to increase the critical temperature significantly.

We start with Hamiltonian of almost ideal  Fermi and Bose gases:
\begin{equation}
H=H_F+H_B+H_{BF}
\label{hamiltonian}
\end{equation}
with
\begin{eqnarray*}
\displaystyle H_F &=& \sum_{\alpha  p} \xi_{p } a_{\alpha  p
}^{\dagger} a_{\alpha p }  
+ \frac{1}{2} U_{FF}\sum_{\alpha \beta   p p' q} a_{\alpha   p-q
}^{\dagger} a_{ \beta  p'+q }^{\dagger}
 a_{\beta  p' }  a_{\alpha  p } \\
H_B &=&  \sum _{ \, p} \varepsilon_p b_{ p }^{\dagger} b_{p
} 
+
\frac{1}{2}U_{BB} \sum_{\, p p' q} b_{ p-q }^{\dagger} b_{p'+q }^{\dagger}
 b_{p' }  b_{p } \\
H_{BF} & = & \frac{1}{2}U_{BF} \sum_{ \, p p' q} 
a_{\alpha p-q }^{\dagger} b_{ p'+q }^{\dagger}
 b_{p' }  a_{\alpha p },
\end{eqnarray*}
where $\xi_{\alpha  p } = \left( p^2 / 2m_{ F } - \varepsilon_{F \alpha }
 \right)$, 
$\varepsilon_p = p^2 / 2m_B $ are the  kinetic energies of the 
Fermi and Bose gases
respectively, $\varepsilon_{F \alpha}$ is the Fermi energy of the Fermi-gas
with spin (hyperfine component) $\alpha$,
$U_{FF}$, $U_{FB}$, and $U_{BB}$ are two particle interaction 
constants for Fermi-Fermi, Fermi-Bose and Bose-Bose interactions. The
constants are related to the two-body scattering lengths $a_{BB}$,  $a_{BF}$
and  $a_{FF}$  in vacuo as
$U_{FF}= 4 \pi a_{FF} / m_{F}$, $U_{BB} = 4 \pi a_{BB} / m_{B}$,
$U_{BF}= 4 \pi a_{BF} / m_{BF} $;
$ m_{BF} = 2 m_B m_F / (m_B+m_F)$. 

In general (apart from the case of resonance scattering) the 
harmonics of the scattering amplitude for slow particles are proportional 
to $ f_{\ell} \sim a (a p_F)^{2
\ell }$,  \cite{Landau3}, where $a$ is a length of the order of a $s$-scattering length. 
For example, the $\ell =1$ bare contribution goes like  $U_{FF}^{\ell} \nu_f\sim
(ap_F)^{3}$. This contribution is very small and can in general be 
neglected if some other  triplet pairing mechanism is present. 
In the case of fermions in two spin states for instance,
the indirect interaction by polarisation of the fermions 
in the other spin state  
(Kohn-Luttinger mechanism)
\cite{Kohn},
provides a contribution of order $(a p_F)^2$, and is 
therefore more important than the bare one.

A standard procedure \cite{Landau9} yields the critical temperature for pairing
with given angular momentum $\ell$:
\begin{equation}
\label{eq:tc-exp}
T_{c \ell} = \tilde{\varepsilon}_F \exp \left\{ -\frac{1}{\nu_F |
\tilde{\Gamma}_{\ell}|} \right\},
\end{equation}
with $\tilde{\Gamma}_{\ell} < 0$ being the $\ell$-th spherical harmonic   
of the irreducible vertex,  
$\tilde{\varepsilon}_F$ is some energy parameter of the order of the 
Fermi energy, and $\nu_{F}$ the density of states on the Fermi level.
The real transition corresponds to the angular momentum with the maximum 
allowed temperature. The effective interaction between two Fermi particles 
$\tilde{\Gamma}$ is the sum of the bare one
$U_{FF}$,  the interaction via polarisation of the  
bosonic medium (exchange of density fluctuations) $U_{FBF}$, and possibly that via polarisation of the
other fermionic species $U_{FFF}$ if fermions in 
more than one spin orientation are present.

We assume temperatures much smaller than those
of degeneracy. The correctness of this hypothesis will be confirmed by the
results found. 
The effective interaction of Fermi quasi-particles on mass-surface 
with zero transfer energy is given   by 
$U_{FBF}(q) = U_{BF}^2 \chi(q,\omega =0)$,
where $\chi(q,\omega)$ is the density-density response function for an
almost ideal Bose-gas  at zero temperature \cite{note3}.
Since we are interested in the low density limit of Bose and Fermi-gases  
and $U_{BB} \gg \nu_F U_{BF}^2 \sim (a_{BF} p_F)U_{BF}$,  
we can neglect the renormalization of boson density-density correlation
function due to Bose-Fermi interaction,
and we can write to first order in the gas parameter \cite{Nozieres}: 

\begin{equation}
\label{chi}
\chi(q,\omega)  = 
\frac{ \displaystyle 1 } {\omega^2 - \varepsilon_q^0
(\varepsilon_q^0 +2n_B U^{\mbox{}}_{BB})} \cdot \frac{n_B q^2}{m_B}.
\end{equation}

So the effective interaction of Fermi atoms with zero transfer 
energy reads:

\begin{equation}
\label{ufbffin}
U_{FBF}(q,0) = - \frac{U_{BF}^2}{U_{BB}}
\left( 1+\left(\displaystyle \frac{q}{2m_{B} s}\right)^2 
\right)^{-1},
\end{equation}
where $s = (n_B U_{BB}/m_B)^{1/2}$ is the sound velocity in the boson gas.  
We recall that stability requires $U_{BB}>0$.

A direct calculation of the 
first three partial components of $U_{FBF}$ gives the following
results:
\begin{equation}
\label{Ufbf}
\nu_F U_{FBF}^{\ell} = - \nu_F \frac{U_{BF}^2}{U_{BB}} R_{\ell}(p_F/m_B s),
\end{equation}
with
\begin{eqnarray}
&&R_{0}(x) = 
\frac{\ln (1+x^2)}{x^2},
\nonumber \\
&&R_{1}(x)= 
\frac{2}{x^2}\left[ \left( \frac{1}{x^2} +\frac{1}{2}
\right) \ln(1+x^2) -1
\right], \nonumber \\
&&R_{2}(x)=\frac{6}{x^4} \left[
\left(
\frac{1}{x^2} +1+\frac{x^2}{6}
\right) \ln (1+x^2) - \left(1+\frac{x^2}{2} \right)
\right].\nonumber
\end{eqnarray}
\begin{figure}
\hskip 0.1\columnwidth
\psfig{file=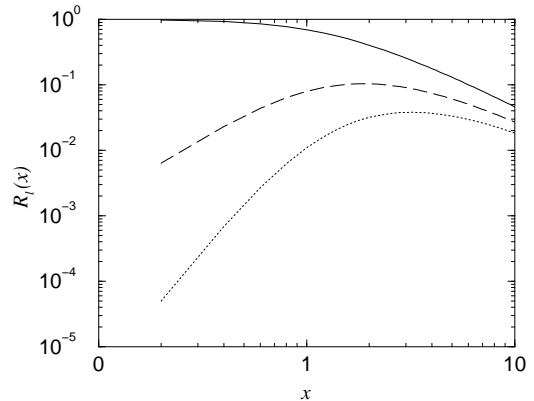,width=0.8 \columnwidth}
\caption{
Functions $R_{0}(x)$ -- solid line; $R_{1}(x)$ -- dotted and
$R_{2}(x)$ -- dashed.}
\end{figure}

The functions $R_{0}(x)$, $R_{1}(x)$ and $R_{2}(x)$ are  shown in Fig 1.
The strongest interaction is in the channel with orbital momentum $\ell = 0$. 
For large $\ell$ one can  show that $R_{\ell}$ drops
off exponentially in $\ell$. Therefore this contribution to the effective 
interaction  for $\ell >2$ can be neglected. 
The functions for $\ell \neq 0$ are strongly non-monotonic in contrast with
zero angular momentum case. For instance, the maximum of $R_1(x)$
is obtained for $x_{opt}=1.86$ ($R_{1}(1.86)=0.1$).
The maximum gives the optimal
ratio of Bose and Fermi-components for given scattering lengths.

Let us consider a binary gas consisting of one fermionic and one bosonic
species. The Cooper pairing in the $s$-wave channel is
prohibited by the Pauli principle. 
It can
exist owing to density fluctuations of bosonic medium with effective
attractive interaction in $p$-wave channel given by formula (\ref{Ufbf}). 
Note that the value of the ratio $U_{BF}^2/U_{BB}$ and the densities
of the gases cannot be arbitrary. The restriction is 
associated with the phase separation at high densities
of the binary mixture into 
two regions: a Fermi--Bose mixture (with densities $n_B^{sep}$ and
$n_{F1}$) and a pure fermionic region (with density $n_{F2}$). 
This phase separation into two large regions is a full analog of that 
observed in  the mixtures of $^3$He--$^4$He.
To check the stability of the mixture against phase separation 
we rewrite  the expression for $U_{FBF}$ in dimensionless
parameters $\lambda$, $\alpha$ and $\beta$ in the 
spirit of ref. \cite{Viverit99}:

\begin{equation} \label{eq:6}
\lambda= \frac{\nu_{F} U_{BF}^2}{U_{BB}} = 
\frac{2}{\pi} \frac{m_{B} m_{F}}{m_{BF}^2} \frac{a_{BF}}{a_{BB}} a_{BF}p_{F} . 
\end{equation}
The expression for $\lambda $ is exactly the factor 
in front of $R_{\ell}$ in (\ref{Ufbf}). 
\begin{equation}
n_B^{sep} = 
\frac{\varepsilon_F}{U_{BF}} (y^2-1) = \frac{(6 \pi^2 n_F)^{2/3} }{8 \pi a_{BF}}\frac{m_{BF}}{m_F}(y^2-1) ,
\end{equation}
where $y \geq 1$ is solution to the equation
\begin{equation}
-\frac{15}{\lambda} (y+1)^2 + 8 y^3 + 16 y^2 +24 y + 12 = 0.
\end{equation}

We then find
\begin{equation}
\label{1ox2}
 \frac{1}{x^2} = \frac{\beta (y(\lambda)^2-1)}{\lambda} ,
\end{equation}
where
$$
\beta =   \frac{\alpha }{ \pi} \cdot \frac{m_B}{m_{BF}} p_F
a_{BF} , \, \, \, \,  \, \,
\alpha = \frac{n_{B}}{n_{B}^{sep}
(\lambda)} . 
$$

The physical meaning of the introduced variables is the following.  
In the case of phase separation  $y$ is the ratio of 
fermionic densities in the two regions $y = n_{F2}/n_{F1} \geq 1$ and $n_B^{sep}$ is
the density of bosonic component in bosonic-fermionic mixture region. 
The problem of phase
separation of binary mixtures of bose and fermi gases was investigated in ref.
\cite{Viverit99}.   
The authors have shown that there are three
possibilities: a) a single uniform phase; b) a purely fermionic phase
coexisting with a mixed phase; c) a purely fermionic phase coexisting with a
purely bosonic.  
Let us examine all three possibilities.

\begin{figure}[t]
\hskip 0.1\columnwidth
\psfig{file=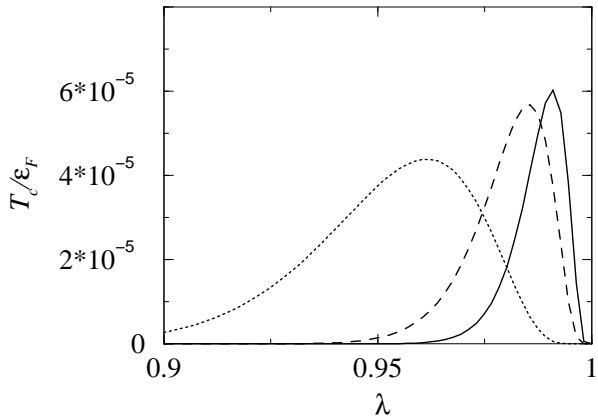,width=0.9\columnwidth}
\caption{$T_c/\varepsilon_F$ versus $\lambda$ for binary boson-fermion
mixture for different values of the coefficient $\beta$. 
The solid curve is for  $\beta=5$, the dashed one for $\beta=3$, 
and the dotted one for $\beta=1$.}
\label{fig:tc}
\end{figure}

The single uniform phase is stable provided the  conditions 
$\lambda \leq 1$ and
$\alpha \leq 1$ are fulfilled. This immediately gives 
$\nu_F U^{2}_{BF} /U_{BB}<1 $ and the
value of the effective interaction is restricted by $U_{FBF}<0.1$, which
corresponds to temperatures of Cooper pairing in the binary mixtures 
about 5 orders of magnitude less than the Fermi energy. 
In figure 2 we plot the critical
temperature as a function of $\lambda$ for given $\beta$. We see, that the
maximum of the critical temperature increases with increasing of $\beta$. 
Note that for given scattering lengths the maximum is in region of the
parameters close to the
phase separation ($\alpha = n_B /n_B^{sep} \to 1 $).   

If the total density of the boson gas is larger than 
$n_B^{sep}$ ($\alpha>1$), the binary mixture
undergoes phase separation into two phases: a purely fermionic phase and the
mixed Fermi-Bose phase. In this case in the mixed phase
the  density of the bosonic gas is $n_{B1} = n_B^{sep}$ and the density of
fermi 
gas $n_{F1}$.  These are determined by the system of the equations
(\ref{eq:6})--(\ref{1ox2}). Our
result obtained for the single uniform phase is still within the mixed phase
valid if the appropriate densities of fermi and bose gases are used. 

The third possibility is the  coexistence of a purely fermionic phase with a
purely bosonic one, which exists for much higher total densities 
of bosons and fermions. In this case of course there
is no effective interaction between fermions due to the exchange of
boson density fluctuations. 

We can conclude that the contribution of the exchange  density 
fluctuations of the bosonic medium has its
maximum for the set of parameters 
close to those of phase separation of a single uniform phase into two
coexisting phases: a mixed phase and a pure fermionic one. 

\begin{figure}[t]
\hskip 0.1\columnwidth
\psfig{file=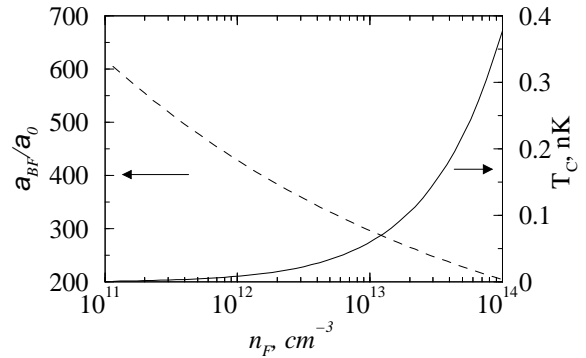,width=0.9\columnwidth,height=0.6\columnwidth}
\caption{ Optimal scattering length $a_{BF}$ and corresponding $T_c$ for Cooper
pairing, versus fermionic density in a binary boson-fermion mixture 
of $^6$Li and $^{87}$Rb.} 
\label{fig:tc-a}
\end{figure}

Let us make some estimates for real systems. Take a binary mixture
of fermionic $^6$Li and bosonic $^{87}$Rb, and choose for $^6$Li the density
$n_F \sim 10^{12}$cm$^{-3}$. This corresponds to 
$\varepsilon_F/k_B \sim 600$nK. 
The boson-boson scattering length is $a_{BB}=110a_0$, $a_0$ being 
the Bohr radius. 
The boson-fermion scattering length is unknown. In order to get $T_c$ close to
its maximum value for binary boson-fermion mixture, we should obtain $\lambda$
close to 0.95. For the given mixture this corresponds to $a_{BF} \sim 450a_0$. 
Substituting these values into (5) -- (9), and taking
for the Bose component the density
$n_B \sim 10^{13}$cm$^{-3}$ we get $T_c \sim 10^{-2}$nK. 
Similar calculations for $n_F \sim 10^{14}$cm$^{-3}$ imply 
$a_{BF} \sim 200 a_0$, and with $n_B \sim 5\times 10^{14}$cm$^{-3}$, 
$T_c \sim 0.5$nK. 
In figure \ref{fig:tc-a} we summerise  the optimal 
parameters of the system and the corresponding critical
temperatures for $p$-wave pairing in a binary mixture.

There is a possibility of increasing
$T_c$ by combining the above mentioned mechanism with either $p$-wave 
quasi-bound 
resonance for the scattering of Fermi-atoms, or by 
considering a mixture of two spin states of fermions with one of bosons. 
In the former case the irreducible
vertex in (\ref{eq:tc-exp}) is determined by the sum of the  
interactions due to
polarisation of bosons and the (now large) 
bare $p$-wave attractive scattering of Fermi-atoms \cite{note1}.

In the case of two species of
fermions with one of bosons, again  
the effective interaction has two contributions:
from boson density fluctuations and
from the  Kohn-Luttinger mechanism.
The effective interaction in the latter
is a non-monotonical function of the ratio of the densities of the  
different hyperfine components (see \cite{Kagan96,Efremov00a}). 
Its maximum is $ \nu_F U_{eff} \sim 0.058 (ap_F)^2$ which corresponds to
a  ratio $n_1 \sim 2.8 n_2$.

For optimal parameters  \cite{Kagan88} and a density $n_{1} =
10^{14}$cm$^{-3}$ the critical temperature reads:
$$
T_c \sim \tilde{\varepsilon}_F \exp{ \left\{
-\frac{1}{\nu_F U_{FBF}+ 0.058 (a_{FF}^s p_F)^{2} } 
\right\}
}.
$$ 

$T_c \sim 1$nK and $T_c \sim 20$nK respectively 
for bare $s$-wave Fermi-Fermi scattering
lengths $|a_{FF}^{s}|  =500a_0$ and $1000a_0$. 
Note that the value of the critical temperature obtained
is valid for $a_{FF}^{s}>0$ as well as for $a_{FF}^{s}<0$. 
Pure Kohn-Luttinger mechanism would give $T_c = 10^{-30} $nK and $10^{-5}$nK 
respectively for the given scattering lengths, so that the main 
contribution comes from the boson-induced term.
For $^6$Li however the $s$-wave scattering length between 
two different hyperfine states
is $a_{FF}^{s}=-2160a_0$. The critical temperature 
with this scattering length
is $ \sim 0.5 \mu$K. For pure Kohn-Luttinger mechanism it would have
been $\sim 0.1\mu$K, which shows the strong effect that the bosons have
also in this case. The full analogy with mixtures $^3$He -$^4$He says that
here also both single uniform phase and phase separated states are possible,
and explicit calculations for the case of
two-Fermi species and a Bose one need to be carried out.

In conclusion we showed that
the fermions in a (typical) dilute binary mixtures of Fermi and Bose gases
are unstable towards $p$-wave Cooper pairing. This is due to their effective
attraction arising from boson polarisation. We then calculated how the 
associated $T_c$ can be maximized.
Although the highest $T_c$'s found don't seem presently experimentally observable,
we showed that the mechanism may be used to enhance pairing when combined with 
others. 

We are very grateful to P. Fulde, A. J. Leggett,  M. Baranov, 
M. Kagan, A. Zvyagin, N. Shannon for useful discussions. 



\begin{thebibliography}{99}
\bibitem{note3} Retardation can in principle have considerable effects
in the limit $p_F /m_B s \sim 1$. 
An analysis similar to that in ref. \cite{Efremov00a} for the $p$-wave
case shows that retardation effects slightly reduce $T_c$. And we assume that
this reduction is absorbed by the prefactor $\tilde{\varepsilon}_F$.
\bibitem{note1} 
For $^6$Li the 
$p$-wave scattering length for atoms in the same hyperfine state
is sufficiently small $a=-35a_0$. However, we can imagine that it is possible
to trap a hyperfine state that has a $p$-wave quasi-bound resonance 
in the spirit of refs. \cite{Fedichev96,Moerdijk96} and tune the parameters of
the system  such  that the $p$-wave scattering length becomes large and
negative.
\end{thebibliography}
  \end{document}